\documentclass[conference]{IEEEtran}
\IEEEoverridecommandlockouts
\usepackage{cite}
\usepackage{amsmath,amssymb,amsfonts}
\usepackage{algorithmic}
\usepackage{graphicx}
\usepackage{textcomp}
\usepackage{xcolor}
\usepackage{multirow}
\usepackage{threeparttable}

\def\BibTeX{{\rm B\kern-.05em{\sc i\kern-.025em b}\kern-.08em
    T\kern-.1667em\lower.7ex\hbox{E}\kern-.125emX}}
\begin{document}

\title{Study on the Particle Sorting Performance for Reactor Monte Carlo Neutron Transport on Apple Unified Memory GPUs
\thanks{}
}

\author{\IEEEauthorblockN{1\textsuperscript{st} Changyuan Liu}
\IEEEauthorblockA{
\textit{New Compute Laboratory}\\
Beijing, China \\
changyuan\_liu@163.com}
}

\maketitle

\begin{abstract}
 In simulation of nuclear reactor physics using the Monte Carlo neutron transport method on GPUs, the sorting of particles plays a significant role in performance of calculation. Traditionally, CPUs and GPUs are separated devices connected at low data transfer rate and high data transfer latency. Emerging computing chips tend to integrate CPUs and GPUs. One example is the Apple silicon chips with unified memory. Such unified memory chips have opened doors for new strategies of collaboration between CPUs and GPUs for Monte Carlo neutron transport. Sorting particle on CPU and transport on GPU is an example of such new strategy, which has been suffering the high CPU-GPU data transfer latency on the traditional devices with separated CPU and GPU. The finding is that for the Apple M2 max chip, sorting on CPU leads to better performance per power than sorting on GPU for the ExaSMR whole core benchmark problems and the HTR-10 high temperature gas reactor fuel pebble problem. The partially sorted particle order has been identified to contribute to the higher performance with CPU sort  than GPU. The in-house code using both CPU and GPU achieves 7.5 times power efficiency that of OpenMC on CPU for ExaSMR whole core benchmark with depleted fuel, and 150 times for HTR-10 fuel pebble benchmark with depleted fuel.
\end{abstract}


\section{Introduction}
Being the method with the highest fidelity, the Monte Carlo method has been adopted as a verification tool to other methods such as discrete ordinates and the method of characteristics. Because of its heavy computation burden, the Monte Carlo method has not been considered as the everyday reactor simulation tool. The great performance improvement on GPUs demonstrated in recent studies makes the adoption of Monte Carlo method as a routine practice more practical. Table~\ref{tab:summary_gpu_mcnt} summarizes some recent work. 
\begin{table}[h]
\centering
\caption{Summary of continuous energy Monte Carlo neutron transport code with GPU support}
\label{tab:summary_gpu_mcnt}       
\begin{tabular}{lll}
\hline
Code & Developer & Sorting on CPUs or GPUs\\
 \hline
Warp~\cite{warp} & Univ. California, Berkeley & CPUs \\
Pragma~\cite{pragma} & Seoul National Univ. &  GPUs\\
Shift~\cite{shift} & Oak Ridge National Lab.  & GPUs \\
OpenMC~\cite{openmc_gpu} & Argonne National Lab.  & CPUs (Possibly) \\
MagiC~\cite{magic} & Univ. South China  & GPU (Possibly) \\
\hline
In-house & In-house & CPUs and GPUs\\
\hline
\end{tabular}
\end{table}

As discovered by Hamilton~\cite{shift}, particle sorting is important for achieving high neutron transport performance by increasing the coherence in execution paths between GPU threads. Joo~\cite{pragma} further elaborates the particle sorting strategies. In previous study, most codes such as Pragma~\cite{pragma}, Shift~\cite{shift} and MagiC~\cite{magic} (possibly) use GPUs for particle sorting, and OpenMC~\cite{openmc_gpu} possibly uses CPUs for particle sorting. The Warp~\cite{warp} code seems sorting particles on CPUs too. 

\begin{figure}[h]
\centering
\includegraphics[width=8cm,clip]{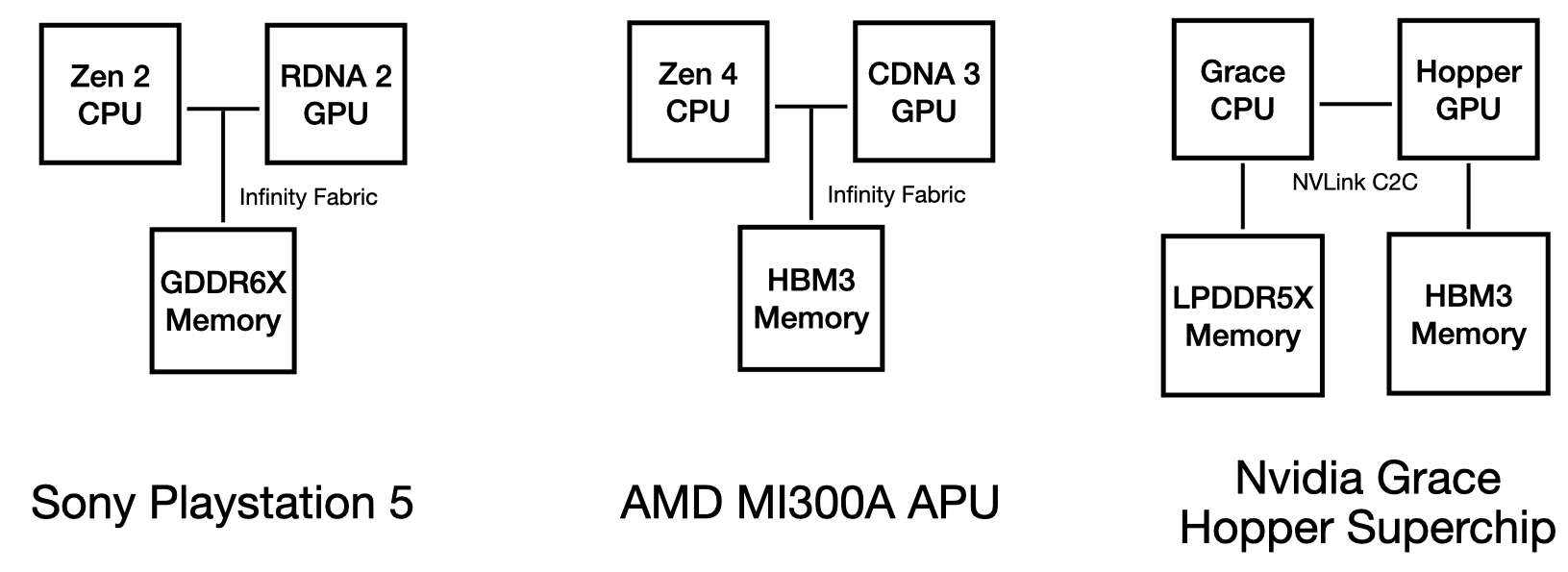}
\caption{A snapshot of the design of some recent merged CPU and GPU chips .}
\label{fig:uma_devices}       
\end{figure}

As indicated in Figure~\ref{fig:uma_devices}, from chips for personal entertainment such as Sony Playstation 5~\cite{ps5} to chips for high performance computation such as AMD~\cite{amd_mi300a} and Nvidia~\cite{nvidia_grace_hopper} have merged CPU and GPU chips. Some chips such as PS5 and MI300A have unified memory with the CPU, GPU and memory connected with the high speed bus called infinity fabric. 

This work proposes to use Apple unified memory computing devices to study the collaboration between CPUs and GPUs in Monte Carlo neutron transport methods. This collaboration is previously uneconomic because of the low data transfer rate and high data transfer latency between CPUs and GPUs on computing devices with separated CPUs and GPUs. There are previous work study the Apple silicon for Monte Carlo methods in areas such as: CPU performance study~\cite{db2022}, multi-core CPU work balance~\cite{apple_balance},  and cross section lookup on GPU~\cite{apple_gpu_lookup}.

The contributions are summarized as followed.
\begin{itemize}
\item Discussion about programming for Apple M2 Max chip
\item Study of the sorting performance on CPU-GPU for partially sorted data
\item Verification of in-house merged CPU-GPU code with VERA pincell and assembly benchmark problems
\item Comparison of CPU and GPU sorting strategies on the simulation power efficiency for ExaSMR whole core and HTR-10 fuel pebble benchmark problems
\end{itemize}

\section{Development on Apple Silicon as a Unified Memory Device}

The Apple silicon chips are system-on-chips (SoCs), where a cluster of more powerful performance CPU cores, and a cluster of less powerful efficiency cores, and a cluster of GPU cores are integrated on the same silicon die. All CPU and GPU clusters have its private L2 cache, and these clusters are sharing an L3 cache named as the System Level Cache (SLC). 

\subsection{Apple M2 Max Chip}

\begin{figure}[h]
\centering
\includegraphics[width=8cm,clip]{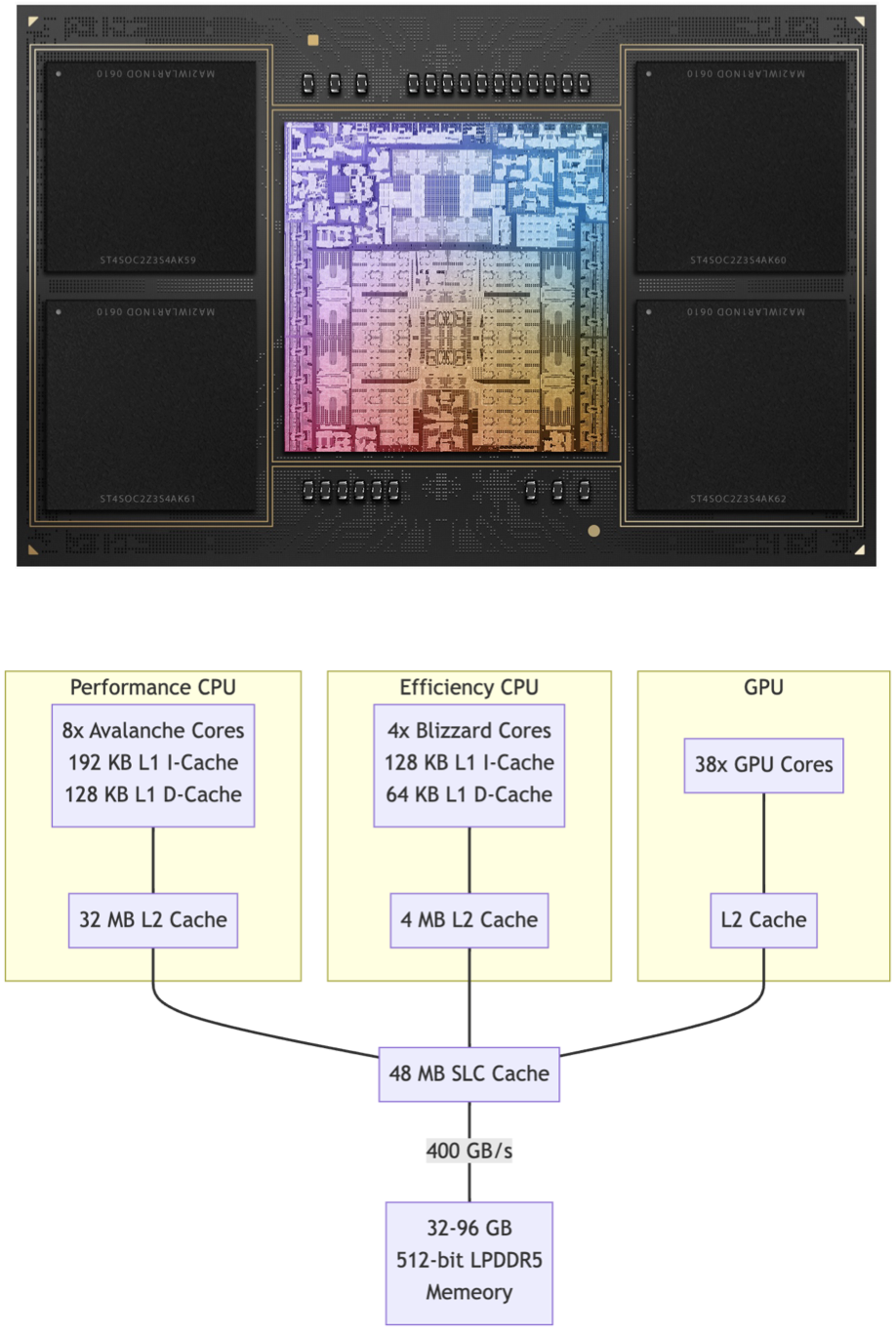}
\caption{A snapshot~\cite{apple_m2max} (left) and a sketch of the design (right) of Apple M2 Max chip. I-Cache stands for instruction cache, and D-Cache stands for data cache. Avalanche and Blizzard are architecture design code names.}
\label{fig:apple_m2max}       
\end{figure}

In this work, the Apple M2 Max chip is studied and Figure~\ref{fig:apple_m2max} gives a snapshot~\cite{apple_m2max} and an illustration of the chip components. There are four memory chips surrounding the SoC in the center.  The memory type is LPDDR5, which offers an interface of 512 bit with a bandwidth of 400 GB/s. In most gaming GPUs, GDDR6 and GDDR6X are the most common types, and in workstation GPUs, HBM2 and HBM3 are the most common types. Usually, LPDDR5 is used for power efficient mobile devices, and LPDDR5 has higher latency. The way of Apple's use of LPDDR5 is unusual, and it has a tight connection with the SoC. The result of this tight packaging is the lower latency than the usual LPDDR5 packaging while keeping the power consumption at low level.


The SoC includes 8 performance CPU cores sharing 32 MB L2 cache and 4 efficiency CPU cores sharing 4 MB L2 cache. The L2 cache is much larger than Intel, AMD and many ARM based CPUs. There are 38 GPU cores sharing an unknown size of L2 cache. Moreover, there is a system level cache (SLC) of 48 MB for all CPU cores and GPU cores. 


What makes the Apple SoC unique is that the CPU and GPU are sharing the same memories and there is a single SLC for both CPU and GPU. Such a design enables closer collaboration between CPUs and GPUs. Table~\ref{tab:comparison_uma} illustrates some of the difference between Apple SoC and systems with discrete GPUs. The close connection between CPU and GPU in Apple SoC enables collaborated CPU-GPU algorithms with frequent communication between CPU and GPU.

\begin{table}[h]
\centering
\caption{Comparison of Apple SoC and systems of CPU with discrete GPU}
\label{tab:comparison_uma}       
\begin{tabular}{lll}
\hline
 & Apple SoC  & Discrete GPU \\\hline
CPU-GPU bus & in-silicon & PCI-E \\
Memory type & sharing & host/device \\
GPU memory latency & low & high \\
\hline
\end{tabular}
\end{table}

\subsection{Objective-C/Swift Programming Languages and Frameworks}
The operating systems MacOS for laptops and workstations, and iPadOS for tablets, and iOS for mobile phones, and watchOS for watches, and tvOS for home media stations, and visionOS for the recently released space computing goggles are delivered with user interfaces with distinguished styles. The basic design of such user interfaces is dated back to the 1980s, where C++ has not yet been prevailing. Another object-oriented language Objective-C ~\cite{objc} inspired from the SmallTalk ~\cite{smalltalk} is adopted by Apple to develop the user interfaces.

\begin{figure}[h]
\centering
\includegraphics[width=7cm,clip]{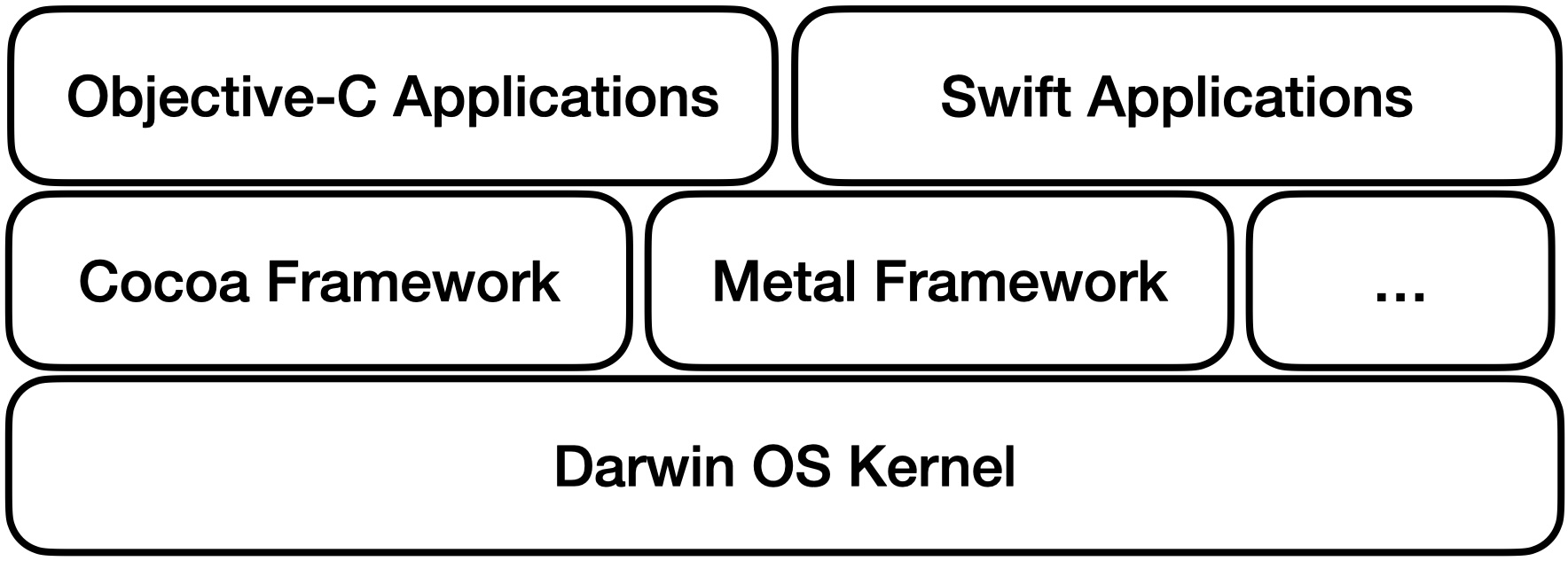}
\caption{A sketch of application development in Objective-C \& Swift programming language on Apple devices.}
\label{fig:apple_frameworks}       
\end{figure}

Later, in the last decade, the Swift ~\cite{swift} language is further proposed for meeting the demand of software developers for an easier to use languages. Applications developed in Objective-C or Swift are integrated with system frameworks such as Cocoa~\cite{cocoa} for user interfaces and Metal~\cite{metal} for 3D graphics. Figure~\ref{fig:apple_frameworks} illustrates the layers of applications, frameworks and OS kernel.

At the lowest level, Apple computing devices run the Darwin OS kernel~\cite{darwin}, which is different from Linux. Same as Linux, Darwin implements the Portable Operating System Interface (POSIX)~\cite{posix}. So migration of lower level applications between Linux and Darwin is much easier than that between Linux and Windows, where the POSIX has not been completely implemented on Windows. As a side notice, Windows has provided the Windows Subsystem for Linux (WSL)~\cite{wsl} to provide an embedded Linux environment, in order to execute Linux application on Windows.

\subsection{Metal Shading Language \& Framework}
At the beginning, Apple did not design its own programming languages for GPUs. Instead, OpenGL ~\cite{opengl} and OpenCL ~\cite{opencl} are adopted, which are open standards conceived by many vendors.

However, as the Apple GPUs get more powerful, the OpenGL and OpenCL have been not able to keep the pace of increased hardware features provided by Apple chips. So, the Metal Shading Language(MSL)~\cite{metal} has been proposed. 

\begin{figure}[h]
\centering
\includegraphics[width=8cm,clip]{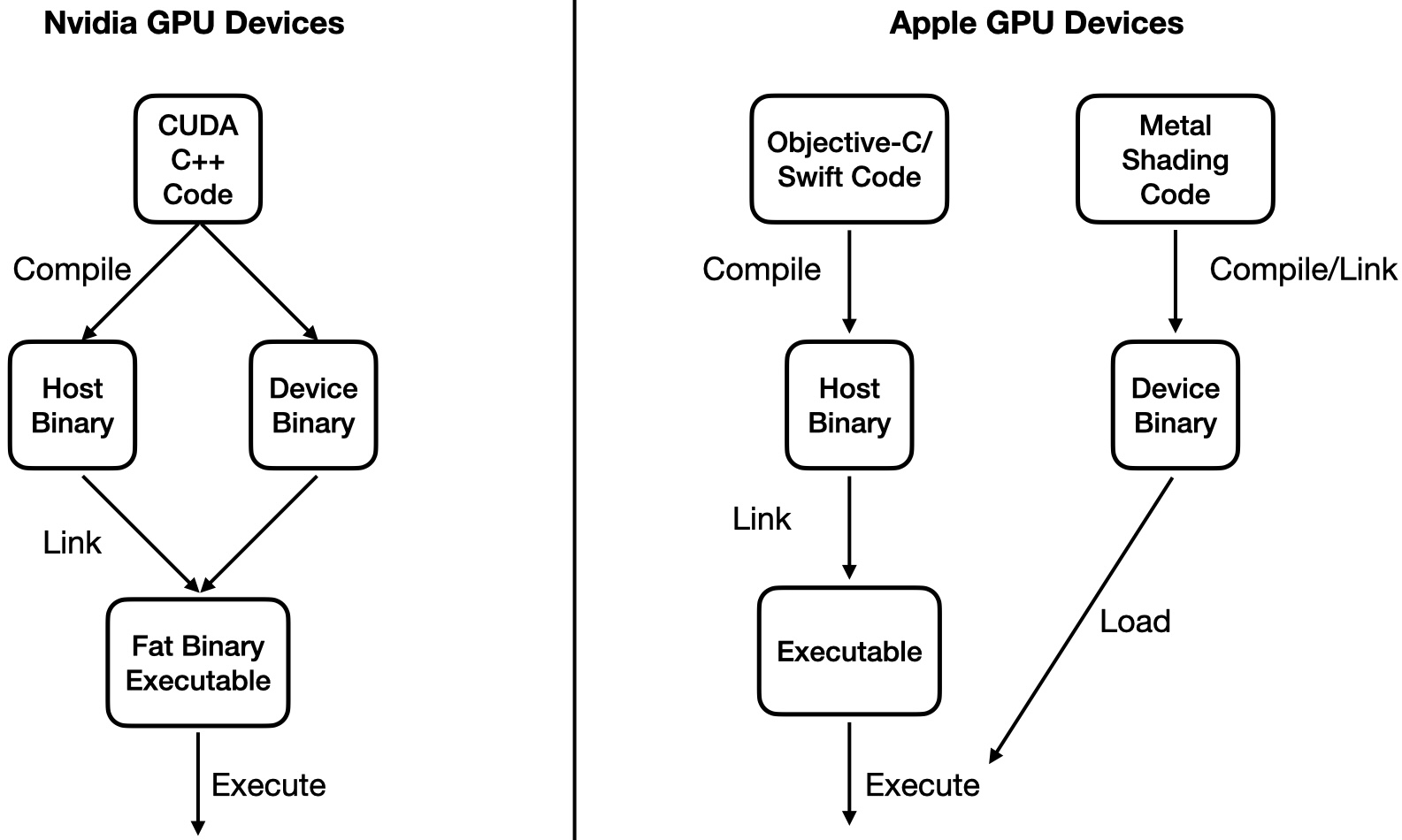}
\caption{A sketch of CPU-GPU program compilation scheme on Nvidia and Apple GPU devices.}
\label{fig:gpu_compilation}       
\end{figure}

Applications written in MSL rely on toolchains provided by the Metal framework for compilation and execution. Although both MSL and CUDA~\cite{cuda}  C++ are based on C++, there are differences in the code building stages. Figure~\ref{fig:gpu_compilation} illustrates the major difference. 

In CUDA, the host code running on CPU and device code running on GPU are combined in the same CUDA C++ source code, while in Metal, the host code in Objective-C or Swift and device code in Metal are separated. Also, in CUDA the CPU and GPU binaries are packed in a single executable, while in Metal, the CPU executable will load Metal GPU code in runtime.

\subsection{Apple GPU Programming Patterns}
\begin{figure}[h]
\centering
\includegraphics[width=6cm,clip]{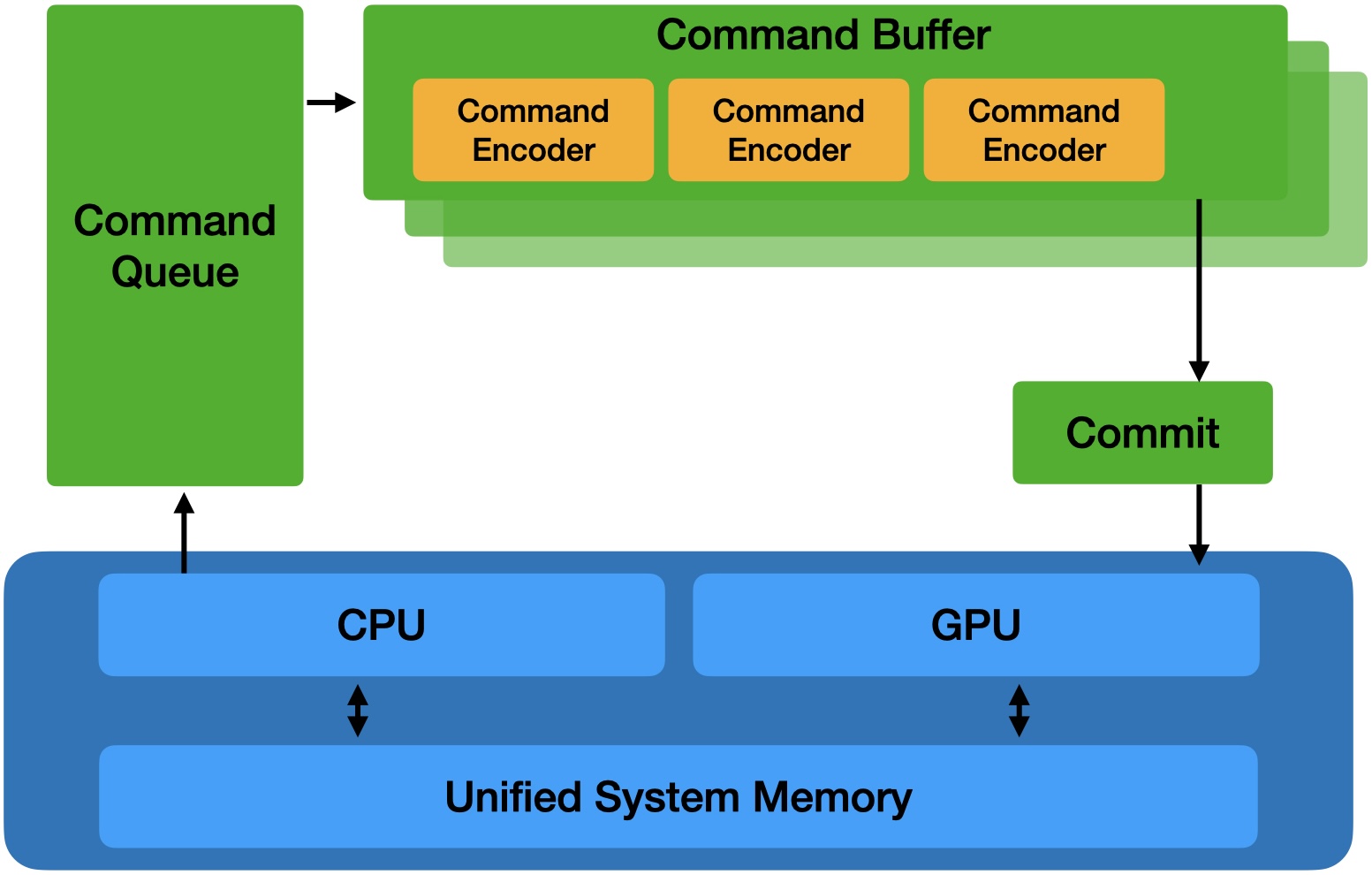}
\caption{Programming patterns for Apple GPU.}
\label{fig:apple_metal_patterns}       
\end{figure}

Programming with the Metal framework on Apple GPU begins with the creation of command queue. Then, create command buffers to submit tasks to GPUs. Each task may contain multiple stages. Each stage creates a command encoder, and each GPU kernel function binds to a command encoder. After all commands in the buffer are encoded, the buffer is committed, so that the GPU starts to execute the commands as soon as possible. Figure~\ref{fig:apple_metal_patterns} illustrates this programming pattern.

\section{Sorting Algorithms}\label{sec:apple_sort}
In this section, the CPU and GPU  sorting algorithms are discussed and studied on Apple chips.

\subsection{Summary of Sorting Algorithms on CPU \& GPU}
There are two sorting codes on CPU, which are the C++ standard library (stdlib) utility and Intel TBB~\cite{tbb} library.  The C++ stdlib adopts the Introsort algorithm and runs on single thread. The average, best and worse case time complexity is $\mathcal{O}(n\log n)$, where $n$ is the number of elements to sort. The Intel TBB library adopts the Quicksort algorithm and supports multi-thread devices. The Quicksort algorithm has the same complexity as Introsort, except that the worse case time complexity is $\mathcal{O}(n^2)$. As a side notice, Introsort is a combination of the three algorthims: Quicksort, Heapsort, and Insertion sort.

Because there are no sorting utilities shipped with the Metal framework, an in-house code has been implemented using the Bitonic sorting algorithm. The average, best and worse case time complexity is $\mathcal{O}(n\log^2 n)$. The Bitonic algorithm requires the data size to be power of 2. Figure~\ref{tab:sorting_algorithms_cpu_gpu} compares the CPU and GPU sorting algorithms.

\begin{table}[h]
\centering
\caption{Sorting algorithms on CPU \& GPU}
\label{tab:sorting_algorithms_cpu_gpu}       
\begin{tabular}{llll}
\hline
 Device & Library & Algorithm   & Time complexity\\
 \hline
CPU & C++ Stdlib (single thread ) & Introsort &  $\mathcal{O}(n\log n)$\\
CPU & Intel TBB (multi-thread)  & Quicksort & $\mathcal{O}(n\log n)$ \\
GPU & In-house & Bitonic & $\mathcal{O}(n\log^2 n)$\\
\hline
\end{tabular}
\end{table}

The time complexity is only a guidance, and the next two subsections propose two experiments to illustrate the performance on Apple chips. 

\subsection{Performance of Sorting on Apple Chip}
\subsubsection{Random Integers}
\begin{figure}[h]
\centering
\includegraphics[width=8cm,clip]{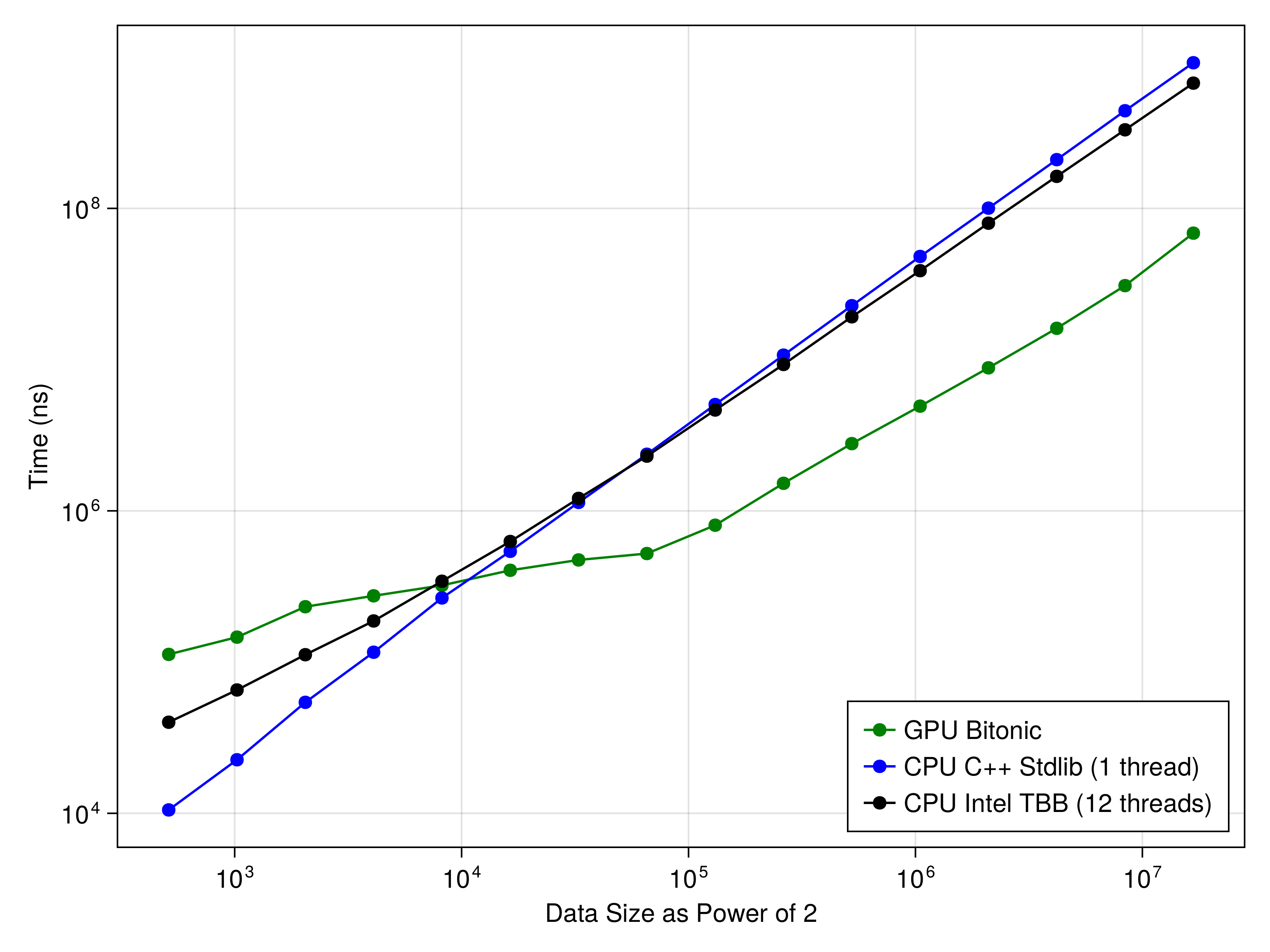}
\caption{Comparison of time cost for sorting integer arrays with size from $2^9$ to $2^{24}$}
\label{fig:random_integers}       
\end{figure}

The first experiment  studies the sorting algorithms on an array of integers randomly sampled. If there are $n$ integers, then each integer is sampled using a uniform distribution between 0 and $n-1$. Figure~\ref{fig:random_integers} compares the time cost for sorting integer arrays with size from $2^9$ to $2^{24}$.

On Log-Log scale, the plot of time cost versus data size appears as straight lines. On GPU, this `straight line' appearance does not extend well below $10^5$. This is because of the GPU execution overhead. Notice that the time measured is purely the GPU execution cost, not including the GPU kernel launch cost.

\subsubsection{Partially Sorted Integers}

It worths notice that the performance of sorting is limited by memory bandwidth. So, for partial sorted data, since there are less data move operations than fully random data, some algorithms may perform better.

\begin{figure}[h]
\centering
\includegraphics[width=8cm,clip]{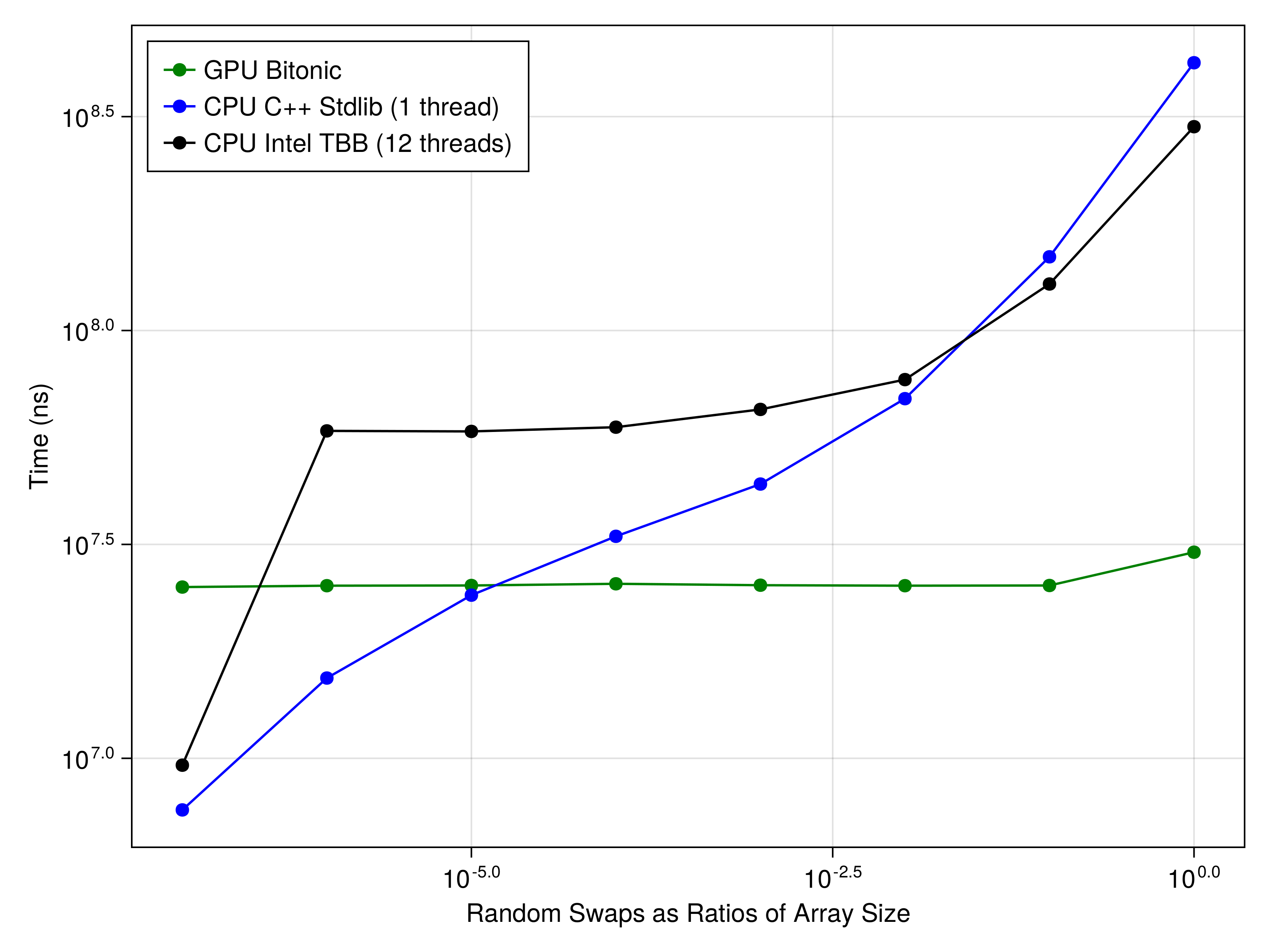}
\caption{Comparison of time cost for sorting integer arrays with size $2^{23}$ and ratio of swaps from 0 to 1.}
\label{fig:partially_sorted_integers}       
\end{figure}

To test the performance of sorting of partially sorted integers, it begins with an array of fully sorted integers. If the are $n$ integers, then the array is $0,1,2,\dots n-1$. Next, define a ratio of swap $r$, and randomly swap $\lfloor nr \rfloor$ pairs of integers in the array, with the pair indices randomly sampled. Here, $\lfloor nr \rfloor$ takes the max integer less or equal to $nr$. Figure~\ref{fig:partially_sorted_integers} shows the time cost for integer arrays of size $2^{23}$ with ratio of swap $r$ from $10^{-7}$ to 1. When $r=10^{-7}$, there are no swaps, so the ratio of swap is essentially 0.

When the number of swaps varies, the GPU Bitonic algorithm performance keeps nearly the same, but the CPU algorithms drastically varies. When there are less than $10^{-5}$ of elements are swapped, CPU performs better than GPU. 

\subsection{Sorting Strategy for Monte Carlo Neutron Transport}
The particle sorting algorithm is important for accelerating Monte Carlo neutron transport on GPU. Hamilton~\cite{shift}, Joo~\cite{pragma}, and Tramm~\cite{openmc_gpu} have good summaries, and Liu~\cite{arxiv2022} discusses the sorting algorithms on Apple computing devices.

\section{Reactor Simulation Benchmarks}
The previous discussion of sorting algorithm on integer arrays is limited, and the results may not reflect the situation of reactor physics simulation. In this section, the VERA pincell and assembly problems~\cite{vera} are simulated to verify the correctness of the program. Then the ExaSMR~\cite{exasmr} whole core and HTR-10~\cite{htr10} fuel pebble benchmark problems are simulated to study the performance.

\subsection{Simulation Configuration}
The in-house code on GPU uses 32-bit floating point number since Apple GPUs only support 32-bit floating point numbers. Instead, OpenMC uses 64-bit floating point numbers.

The cross sections are prepared in an optimal set of 13 temperatures for the kernel reconstruction Doppler broadening method, which is suitable for neutron transport in continuously variable media~\cite{db2022}. For OpenMC, cross sections at the same set of temperatures are used, and the 2-point interpolation Doppler broadening method is used.

The in-house code tallies flux of a 23-group structure and the power. OpenMC code tallies nothing. Table~\ref{tab:configuration} summarizes the simulation configuration. 

\begin{table}[h]
\centering
\caption{Simulation configuration for neutron transport}
\label{tab:configuration}       
\begin{tabular}{lll}
\hline
 & In-house Code & OpenMC Code \\\hline
 Floating precision & 32-bit (single) & 64-bit (double)\\
 Unresolved resonance & turned off & turn off \\
 Resonance scattering & turned off & turn off \\
 Thermal scattering $S(\alpha,\beta)$ & turned off & turn off \\
 Cross section temperatures (K) & \multicolumn{2}{c}{300, 304.252, 338.681, 412.408, 530.512, 705.793,} \\
 & \multicolumn{2}{c}{951.89, 1283.538, 1704.703, 2189.43, 2653.095, } \\
 & \multicolumn{2}{c}{2950, 3000} \\
 Doppler broadening & kernel reconstruction & 2-point linear interpolation\\
 Tally & 23-group flux and power & None\\
 Nuclear data library & ENDF/B-VIII.0 & ENDF/B-VIII.0 \\
\hline
\end{tabular}
\end{table}

\subsection{Verification: VERA Pincell \& Assembly Benchmark Problem}
In order to verify simulation on Apple GPU, the VERA pincell and assembly benchmark problems are studied. Table~\ref{tab:vera_pincell_assembly} compares K-effective values between in-house code on Apple M2 Max CPU+GPU and OpenMC code on Apple M2 Max CPU. The probability table, thermal scattering, and resonance scattering are not considered. There are 1,048,576 particles per cycle with 100 inactive cycles and 200 total cycles.
\begin{table}[h]
\centering
\caption{K-effective of VERA pincell  assembly benchmark problems}
\label{tab:vera_pincell_assembly}       
\begin{tabular}{llllll}
\hline
 & In-house & OpenMC & & In-house & OpenMC\\
& CPU+GPU & CPU only & & CPU+GPU & CPU only \\\hline
1A & 1.18705 (8) & 1.18805 (8) & 2E & 1.06910 (7) & 1.06995 (9)  \\
1B & 1.18190 (9) & 1.18290 (10) & 2F & 0.97484 (8) & 0.97557 (8) \\
1C & 1.17186 (9) & 1.17257 (9) & 2G & 0.84713 (6) & 0.84804 (9) \\
1D & 1.16345 (10) & 1.16405 (9) & 2H & 0.78723 (7) & 0.78799 (8) \\
1E & 0.77405 (7) & 0.77529 (6) & 2I & 1.18092 (8) & 1.18178 (8) \\
2A & 1.18315 (8) & 1.18391 (8) & 2J & 0.97392 (8) & 0.97481 (8) \\
2B & 1.18398 (8) & 1.18471 (8) & 2K & 1.02330 (8) & 1.02385 (8) \\
2C & 1.17466 (8) & 1.17532 (9) & 2L & 1.02126 (7) &  1.02146 (9) \\
2D & 1.16689 (8) & 1.16772 (8) & 2M & 0.94233 (6) & 0.94209 (9) \\
\hline
\end{tabular}
\end{table}

The GPU code underestimates the K-effective within 100 pcm, and the using of single precision floating point numbers play an important role in this discrepancy.

\subsection{Performance Study: ExaSMR Whole Core Benchmark Problem}
Next, the influence of the sorting on the performance of whole core nuclear reactor simulation has been studied with the ExaSMR benchmark problems. Table~\ref{tab:exasmr_complexity} summarizes these problems. There are two versions, one contains fresh fuel with only 7 nuclides in fuel, and the other one contains depleted fuel with 245 nuclides in fuel.
\begin{table}[h]
\centering
\caption{Summary of ExaSMR whole core benchmark simulation}
\label{tab:exasmr_complexity}       
\begin{tabular}{lll}
\hline
 & Fresh fuel & Depleted fuel  \\\hline
Number of nuclides & 76 & 283 \\
Number of nuclides in fuel & 7 & 245 \\
Number of cycles & \multicolumn{2}{c}{350} \\
Number of inactive cycles & \multicolumn{2}{c}{100} \\ 
OpenMC particles per cycle & \multicolumn{2}{c}{1,048,576 ($2^{20}$)} \\
In-house code particles per cycle & \multicolumn{2}{c}{8,388,608 ($2^{23}$)} \\
OpenMC tally & \multicolumn{2}{c}{None} \\
In-house code tally & \multicolumn{2}{c}{fission power + 23-group fluxes} \\
\hline 
K-effective OpenMC CPU only & 1.00656 (6) & 1.00660 (5) \\
K-effective In-house CPU+GPU & 1.00587 (2) & 1.00586 (2) \\
\hline
\end{tabular}
\end{table}

\begin{figure}[h]
\centering
\includegraphics[width=6cm,clip]{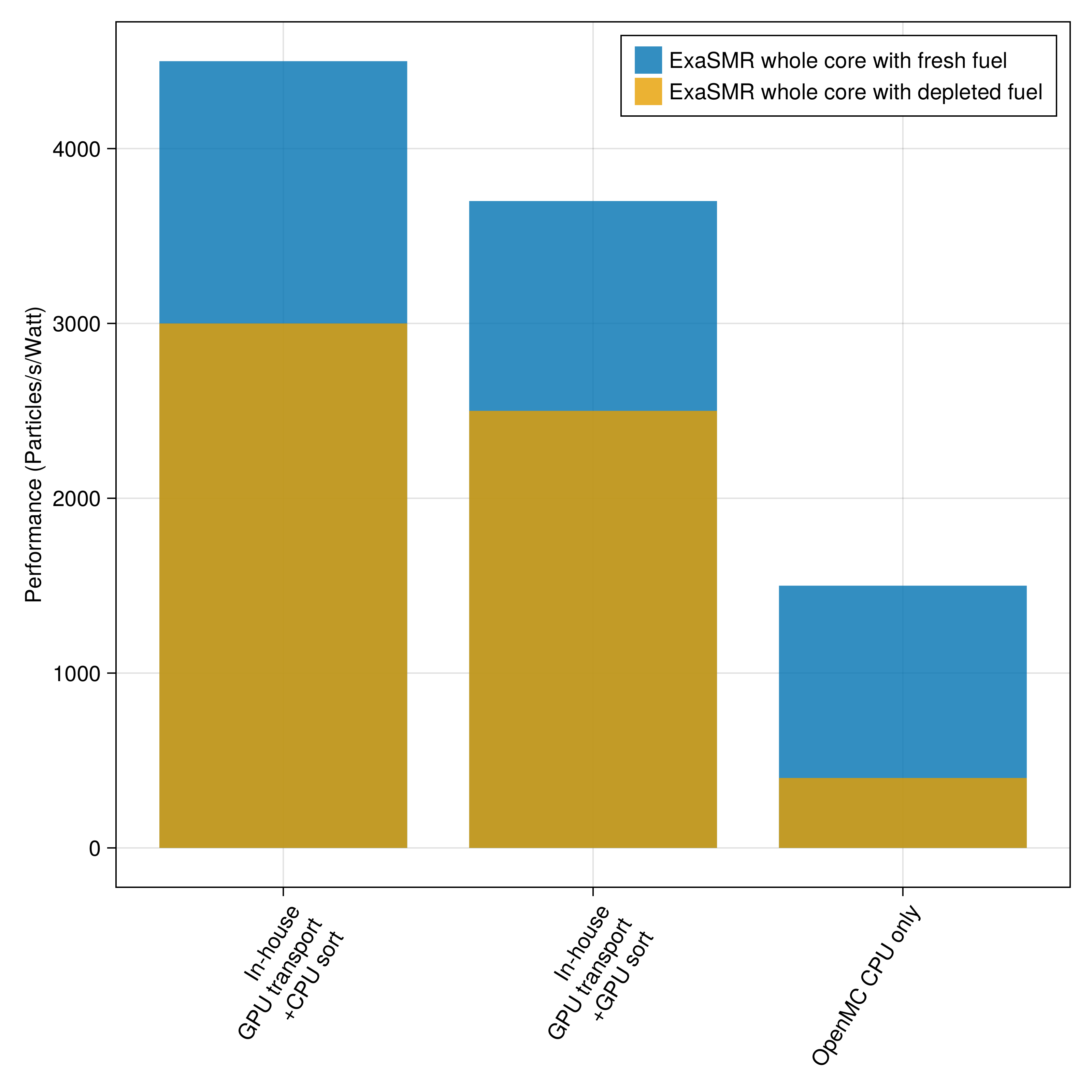}
\caption{Comparison of simulation efficiency in particle per second per Watt for the ExaSMR whole core benchmark problem.}
\label{fig:smr_performance}       
\end{figure}

The simulation performance is summarized in Table~\ref{tab:exasmr_results}. The sorting on CPU performs better than sorting on GPU. This attributes to the partially sorted order in the particles as studied in Section~\ref{sec:apple_sort}. For the fresh fuel problem, the in-house code with GPU transport achieves about 3.0 times power efficiency that of OpenMC, and about 7.5 times for the depleted fuel problem. The power efficiency has been visualized in Figure~\ref{fig:smr_performance}. 

\begin{table}[h]
\centering
\caption{Performance of sorting for ExaSMR whole core benchmark  problems}
\label{tab:exasmr_results}       
\begin{tabular}{llll}
\hline
& In-house & In-house & OpenMC \\
 & sorting on CPU & sorting on GPU & \\
 & active cycles & active cycles & active cycles \\
 & (particles/s/Watt) & (particles/s/Watt) & (particles/s/Watt) \\
\hline
Fresh fuel & 4.5E3 & 3.7E3 & 1.5E3 \\
Depleted fuel & 3.0E3 & 2.5E3 & 4.0E2 \\ 
\hline 
\end{tabular}
\end{table}

\subsection{Performance Study: Pebble Fuel from HTR-10 Test Reactor}
In order to verify the influence of sorting algorithms on the performance of simulation of emerging high temperature gas reactors, the fuel pebble benchmark problem of the HTR-10 test reactor has been studied. High temperature gas reactors have distinguished design from the light water reactors, and the previous study of the performance with ExaSMR may not apply to HTR-10. The definition of the HTR-10 pebble benchmark problem and the simulation configuration and calculated K-effective is summarized in Table~\ref{tab:htr10_complexity}. The code simulation configuration follows Table~\ref{tab:configuration}. The material for the HTR-10 pebble problem is given in Appendix~\ref{sec:htr10_materials}. As indicated in Table~\ref{tab:configuration}, ENDF/B-VIII.0 is used.

\begin{table}[h]
\centering
\caption{Summary of HTR-10 benchmark simulation}
\label{tab:htr10_complexity}       
\begin{threeparttable}
\begin{tabular}{lcc}
\hline
 & \multicolumn{2}{c}{HTR-10 fuel pebble}  \\
 & Fresh fuel & Depleted fuel \\
 \hline
Pebble/fuel region radius (cm) &  \multicolumn{2}{c}{3.0/2.5}\\
TRISO particles in fuel region & \multicolumn{2}{c}{8,335} \vspace{1ex}\\
TRISO fuel/buffer/PyC1/SiC/PyC2  &  \multicolumn{2}{c}{0.025/0.034/0.038/0.0415/0.0455} \\
layers outer radius (cm) & & \vspace{1ex}\\
Pebble boundary condition & \multicolumn{2}{c}{Reflected} \\
Number of nuclides & 10 & 248 \\
Number of nuclides in fuel & 5 & 243 \\
Temperature & \multicolumn{2}{c}{300K} \\
OpenMC particles per cycle & 262,144 ($2^{18}$) & 65,536 ($2^{16}$) \\
In-house code particles per cycle & \multicolumn{2}{c}{1,048,576 ($2^{20}$)} \\
OpenMC tally & \multicolumn{2}{c}{None} \\
In-house code tally & \multicolumn{2}{c}{None} \\
\hline 
K-effective OpenMC CPU only & 1.68732 (9) & 1.68714 (21) \\
K-effective In-house CPU+GPU & 1.68728 (5) & 1.68726 (5) \\
Reference~\cite{triso_uncertainty_previous}\tnote{*} & 1.70534 (13) & N/A \\
\hline
\end{tabular}
\begin{tablenotes}
\footnotesize
\item[*]Assume room temperature with unresolved resonance considered
\end{tablenotes}
\end{threeparttable}
\end{table}

The simulation performance of both in-house code using CPU and GPU sorting and the OpenMC code on CPU is summarized in Table~\ref{tab:htr10_results}. The in-house code agrees with OpenMC in terms of K-effective within in 30 pcm (consider 3 standard deviations). The K-effective is about 500 pcm lower than the reference~\cite{triso_uncertainty_previous}, which is possibly attributed to the ignorance of unresolved resonance (URR).

Same as the ExaSMR whole benchmark, the CPU sorting algorithms perform better than GPU sorting on the basis of performance per power. And the in-house code is about 270 times more power efficient than OpenMC on CPU for pebble with fresh fuel, and 150 times for pebble with depleted fuel.
\begin{table}[h]
\centering
\caption{Performance of sorting for HTR-10 benchmark  problems}
\label{tab:htr10_results}       
\begin{tabular}{llll}
\hline
& In-house & In-house & OpenMC \\
 & sorting & sorting & \\
 & on CPU & on GPU & \\
 & \multicolumn{3}{c}{(particles/s/Watt)} \\
\hline
HTR-10 fuel pebble & 1.7E3 & 1.3E3  & 6.4\\
with fresh fuel & & & \vspace{1ex}\\
HTR-10 fuel pebble & 8.6E2 & 7.7E2 & 5.8\\
with depleted fuel & & & \\
\hline 
\end{tabular}
\end{table}

\section{Conclusions}
In this work, the influence of particle sorting algorithms on the VERA pin and assembly, ExaSMR whole core, and HTR-10 fuel pebble benchmark problems have been studied with the Apple unified memory chips with merged CPU and GPU. First, it has reviewed the programming details on Apple silicon chips. Second, it has demonstrated that with partially sorted data, sorting on Apple M2 Max CPU can outperform GPU. Third, it has verified the correctness of the in-house CPU-GPU code with VERA pincell and assembly benchmarks. Fourth, it has given evidence that the CPU sort is more efficient in power for the ExaSMR whole core benchmark than GPU sort, and the in-house CPU-GPU code achieve 3.0 and 7.5 times power efficiency that of OpenMC CPU code for the case of fresh and depleted fuel. And finally, it has shown that the CPU sort is more efficient in performance per power than GPU sort for the HTR-10 fuel pebble benchmark problem, and the in-house CPU-GPU code achieves 270 and 150 times power efficiency that of OpenMC CPU code for the case of fresh and depleted fuel. In the future, when unified memory  chips with merged CPU and GPU are prevailing, CPU and GPU collaboration methods might be considered for Monte Carlo reactor neutron transport method with better power efficiency.

\section*{Acknowledgment}

Computing technologies from New Compute Laboratory are used to produce parts of the data in this article. New Compute  Laboratory \& its information providers endeavor to ensure the accuracy \& reliability of the information provided, but do not guarantee completeness or reliability, or that it is up-to-date \& accepts no liability (whether in tort or contract or otherwise) for any loss or damage, whether direct or indirect, arising from errors, inaccuracies or omissions or the information being up-to-date. Any information provided is at the user’s risk.

\appendix
\section{Materials for HTR-10 Pebble Benchmark}\label{sec:htr10_materials}
The definitions of materials used in the HTR-10 pebble benchmark are summarized in Table~\ref{tab:htr10_materials}.
\begin{table}[h]
\centering
\caption{Definitions of materials in the HTR-10 pebble problem with fresh fuel}
\label{tab:htr10_materials}       
\begin{tabular}{lll}
\hline
 Material & Nuclide  & Atomic density ($10^{24}$cm$^{-3}$) \\
 \hline
Pebble Carbon matrix & B-10 & 2.49298E-8 \\
& B-11 & 1.00345E-7 \\
& C-12 & 8.57768E-2 \\
& C-13 & 9.60880E-4 \\
\hline
Fuel kernel & B-10 & 4.06384E-7 \\
& B-11 & 1.63575E-6 \\
& O-16 & 4.64720E-2 \\
& U-235 & 3.99198E-3 \\
& U-238 & 1.92441E-2 \\
\hline
Buffer & B-10 & 1.58513E-8 \\
& B-11 & 6.38035E-8 \\
& C-12 & 5.45401E-2 \\
& C-13 & 6.10964E-4 \\
\hline
Pyrolytic Carbon (PyC) & B-10 & 2.73795E-8 \\
inner and outer & B-11 & 1.10206E-7 \\
& C-12 & 9.42057E-2 \\
& C-13 & 1.05530E-3 \\
\hline
Silicon Carbide (SiC) & C-12 & 4.72306E-2 \\
& C-13 & 5.29082E-4 \\
& Si-28 & 4.40486E-2 \\
& Si-29 & 2.23666E-3 \\
& Si-30 & 1.47442E-3\\
\hline
\end{tabular}
\end{table}

\end{document}